\pdfoutput=1
\documentclass[letterpaper,twocolumn,10pt]{article}
\usepackage{usenix,epsfig,endnotes}

\graphicspath{{./figures/}}
\DeclareGraphicsExtensions{.pdf,.png}
\usepackage{comment}
\usepackage{lipsum}
\usepackage[cmex10]{amsmath}
\usepackage{color} 
\usepackage[usenames,dvipsnames,svgnames]{xcolor}
\usepackage{microtype}
\usepackage{vwcol}
\usepackage{multirow}
\usepackage{multicol}
\usepackage{pifont}
\usepackage[T1]{fontenc}
\usepackage{subcaption}

\usepackage{listings}
\lstdefinestyle{embedded}{
	numbers=none,
	frame=none,
	xleftmargin=0cm,
	backgroundcolor=\color{Lavender},
	framesep=1pt,
	aboveskip=3pt,
	belowskip=3pt,
}
\lstdefinestyle{small}{
	basicstyle=\linespread{0.9}\footnotesize,
}

\mathchardef\hyphenmathcode=\mathcode`\-

\let\origlstlisting=\lstlisting
\let\endoriglstlisting=\endlstlisting
\renewenvironment{lstlisting}
    {\mathcode`\-=\hyphenmathcode
     \everymath{}\mathsurround=0pt\origlstlisting}
    {\endoriglstlisting}

\usepackage{array}
\usepackage{mdwmath}
\usepackage{mdwtab}
\usepackage{float}
\newfloat{program}{t}{lop}
\usepackage{fixltx2e}
\usepackage{xcolor}
\usepackage{xspace}

\usepackage{url}
\usepackage{breakurl}
\expandafter\def\expandafter\UrlBreaks\expandafter{\UrlBreaks\do-\do_}

\usepackage[implicit=false, allbordercolors=SkyBlue]{hyperref}

\usepackage{balance}

\usepackage{makecell}
\usepackage{include/grumble}

\newcommand{\myparagraph}[1]{\smallskip \noindent{\bf {#1}.}}

\newcommand{\code}[1]{\mbox{\texttt{#1}}}

\newcommand{\secref}[1]{$\S$\ref{sec:#1}}

\newcommand{\figref}[1]{Figure~\ref{fig:#1}}

\newcommand{\lfence}{\code{LFENCE}}
\newcommand{\bcb}{Bounds Check Bypass}

\begin{document}

\date{}

\title{\Large \bf You Shall Not Bypass: \\ Employing data dependencies to prevent Bounds Check Bypass \\
\vspace{0.3cm}
\large\normalfont Technical Report \\
\normalsize Revision 2.0 (October 1, 2018)}

\author{
{\rm Oleksii Oleksenko$^\dag$, Bohdan Trach$^\dag$, Tobias Reiher$^\dag$, Mark Silberstein$^\ddag$, and Christof Fetzer$^\dag$}\\
{$^\dag$TU Dresden, $^\ddag$ Technion}
} 

\maketitle



\subsection*{Abstract}

A recent discovery of a new class of microarchitectural attacks called Spectre picked up the attention of the security community as these attacks can circumvent many traditional mechanisms of defense.
One of the attacks---Bounds Check Bypass---can neither be efficiently solved on system nor architectural levels and requires changes in the application itself.
So far, the proposed mitigations involved serialization, which reduces the usage of CPU resources and causes high overheads.
In this report, we explore methods of delaying the vulnerable instructions without complete serialization.
We discuss several ways of achieving it and compare them with Speculative Load Hardening, an existing solution based on a similar idea.
The solutions of this type cause 60\% overhead across Phoenix benchmark suite, which compares favorably to the full serialization causing 440\% slowdown.

\section{Introduction}
\label{sec:introduction}

In 2017, multiple research groups independently discovered a new class of microarchitectural attacks, later called Spectre~\cite{Kocher18}.
These attacks target \emph{speculative execution}, a feature of modern processors that improves CPU utilization by executing certain code paths speculatively, before CPU knows, which of the paths is correct.
For example, if an application has a conditional jump, the CPU could start executing one of the branches before it knows the value of the condition.
It may later find out that the prediction was wrong, at which point it will discard the computed results.
However, the CPU will not discard the changes in the \emph{microarchitectural state}, including the cached data.
The Spectre attacks take advantage of this property to circumvent the existing protection mechanisms and leak secret data.

The original Spectre paper~\cite{Kocher18} described two attacks: Bounds Check Bypass (BCB) and Branch Target Injection (BTI).
While the later attack has been patched by a microcode update, the former stays unfixed.
BCB is viewed as an application vulnerability and Intel explicitly states that its mitigation is a responsibility of software developers~\cite{Intel18a}.

To mitigate Bounds Check Bypass, Intel official guidelines~\cite{Intel18a} suggest using the \lfence{} instruction as an explicit serialization point.
The application developer has to identify vulnerable parts of the application and manually harden them with \lfence{}s to prevent speculation.
However, as demonstrated by the long history of memory errors, vulnerabilities can stay unnoticed (and unpatched) for a long time.

To eliminate possibilities for an attack, we have to protect the entire application.
A naive way of doing so would be to add \lfence{}s after every conditional branch.
Although being effective from the security standpoint, this approach significantly reduces the CPU utilization and causes high overheads.
Our experiments show a runtime overhead of up to 440\% on Phoenix benchmarks~\cite{Phoenix}.

\begin{figure}[t]
    \hspace{0.1cm}
       \begin{minipage}[t]{0.45\columnwidth}
        \begin{lstlisting}[frame=tb,framesep=0pt,aboveskip=10pt,belowskip=0pt,label=algo:txexample]
i = input[0];
if (i < size) {
    secret = foo[i];

    baz = bar[secret]; }
\end{lstlisting}
    \subcaption{Example in C}
    \end{minipage}
        \begin{minipage}[t]{0.45\columnwidth}
        \begin{lstlisting}[frame=tb,framesep=0pt,aboveskip=10pt,belowskip=0pt,numbers=none,label=algo:txexample]
a1 = f0(input)
if (condition)
    secret = read(a1)
    a2 = f2(secret)
    read_or_write(a2)
\end{lstlisting}
    \subcaption{Generalized code pattern}
    \end{minipage}
    \caption{Code snippets vulnerable to Bounds Check Bypass.}
    \label{fig:bcb_pattern}
\end{figure}

In this report, we discuss and evaluate approaches to preventing speculation on a more fine-grained level: using data dependencies.

\begin{figure*}[t]
    \hspace{0.1cm}
    \begin{minipage}[t]{0.4\columnwidth}
        \begin{lstlisting}[frame=tb,framesep=0pt,aboveskip=10pt,belowskip=0pt,label=algo:txexample]
i = input[0];


if (i < 42) {



    address = i * 8;
    secret = *address;


    baz = 100;
    baz += *secret;} \end{lstlisting}
        \subcaption{Vulnerable code}
    \end{minipage}
    \begin{minipage}[t]{0.4\columnwidth}
        \begin{lstlisting}[frame=tb,framesep=0pt,aboveskip=10pt,belowskip=0pt,numbers=none,label=algo:txexample]
i = input[0];


if (i < 42) {
    LFENCE;


    address = i * 8;
    secret = *address;


    baz = 100;
    baz += *secret;}\end{lstlisting}
        \subcaption{\lfence{}-based \newline serialization}
    \end{minipage}
        \begin{minipage}[t]{0.4\columnwidth}
        \begin{lstlisting}[frame=tb,framesep=0pt,aboveskip=10pt,belowskip=0pt,numbers=none,label=algo:txexample]
i = input[0];

PUSH rax;
if (i < 42) {
    LAHF;
    XOR rax, r15;
    POP rax;
    address = i * 8;
    secret = *address;
    XOR r15, secret;
    XOR r15, secret;
    baz = 100;
    baz += *secret;}\end{lstlisting}
        \subcaption{\code{LAHF}-based \newline data dependency}
    \end{minipage}
    \begin{minipage}[t]{0.4\columnwidth}
        \begin{lstlisting}[frame=tb,framesep=0pt,aboveskip=10pt,belowskip=0pt,numbers=none,label=algo:txexample]
i = input[0];
all_ones = 0xFFFF...;
mask = all_ones;
if (i < 42) {
    CMOVGE 0, mask;


    address = i * 8;
    secret = *address;
    secret &= mask;

    baz = 100;
    baz += *secret;} \end{lstlisting}
        \subcaption{Speculative \newline load hardening}
    \end{minipage}
    \begin{minipage}[t]{0.4\columnwidth}
        \begin{lstlisting}[frame=tb,framesep=0pt,aboveskip=10pt,belowskip=0pt,numbers=none,label=algo:txexample]
i = input[0];

XOR i, r15;
if (i < 42) {



address = i * 8;
secret = *address;
XOR r15, secret;
XOR r15, secret;
baz = 100;
baz += *secret;} \end{lstlisting}
        \subcaption{Dependency on arguments}
    \end{minipage}

    \caption{Approaches to preventing Bounds Check Bypass. Listings (c) and (e) assume that R15 is exclusively reserved for creating data dependencies.}
    \label{fig:prevent}
\end{figure*}

\section{Bounds Check Bypass}
\label{sec:background}

In essence, Bounds Check Bypass is a buffer over- or under-read that succeeds even in the presence of traditional protection measures.

Consider the code snippet in \figref{bcb_pattern}a:
Without the bounds check on line 2, an adversary with control over the input can force the load on line 3 to read from any address, including those beyond the array \code{foo}.
Traditionally, vulnerabilities of this type were mitigated with bounds checks (such as the one on line 2) that permit the access only if the address is within the object bounds.

However, there is an issue with this approach.
The underlying assumption of bounds checking is that the instructions run in order, which is not the case in modern pipelined CPUs with branch prediction.
Such a CPU can run the check in parallel with the vulnerable load if it predicts that the check is not likely to fail.
Later, it will find out that the prediction was wrong and discard the speculated load, but, as Spectre~\cite{Kocher18} has proven, its cache traces will stay.
The adversary can access the traces by launching a side-channel attack~\cite{Tromer10,Yarom14}.

Since side-channel attacks only reveal the accessed address and not the loaded value, the load on line 3 is not sufficient.
In our example, only the second load (line 5) will leak the secret:
The adversary will observe an access to \code{bar[secret]} after which deriving the value of \code{secret} is only a matter of subtraction.

In summary, the vulnerable pattern (see \figref{bcb_pattern}b) consists of an adversary-controlled load (line 3) followed by a memory access (line 5) based on the loaded value.
A runtime check (line 2) must protect the load;
otherwise, the pattern turns into a traditional buffer overflow.

\alex{Data dependency? Reordering? Speculation?}


\begin{figure*}[t]
    \centering
    \includegraphics[scale=0.65]{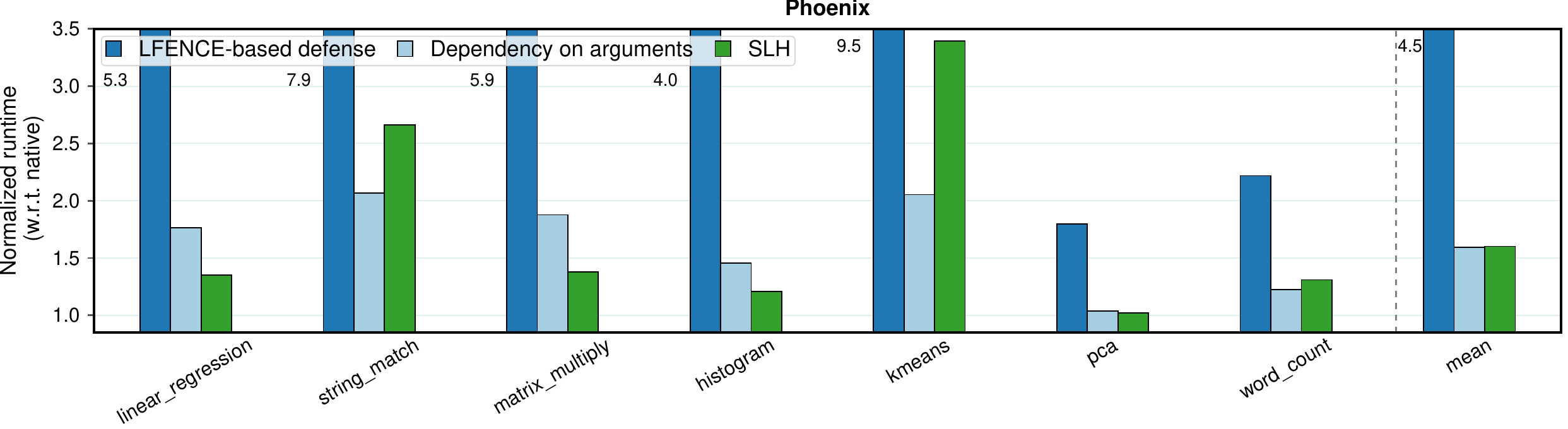}
    \caption{Performance (runtime) overhead with respect to native version. (Lower is better.)}
    \label{fig:perf}
\end{figure*}

\begin{figure*}[t]
    \centering
    \includegraphics[scale=0.65]{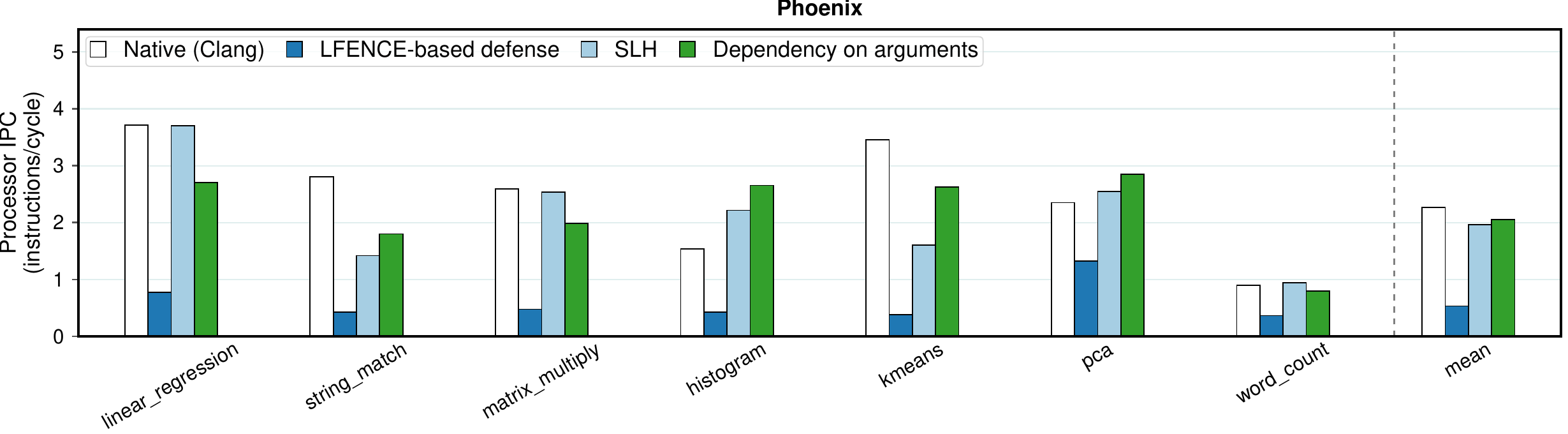}
    \caption{IPC (instructions/cycle) numbers for native and protected versions. (Higher is better.)}
    \label{fig:ipc}
\end{figure*}

\begin{figure*}[t]
    \centering
    \includegraphics[scale=0.65]{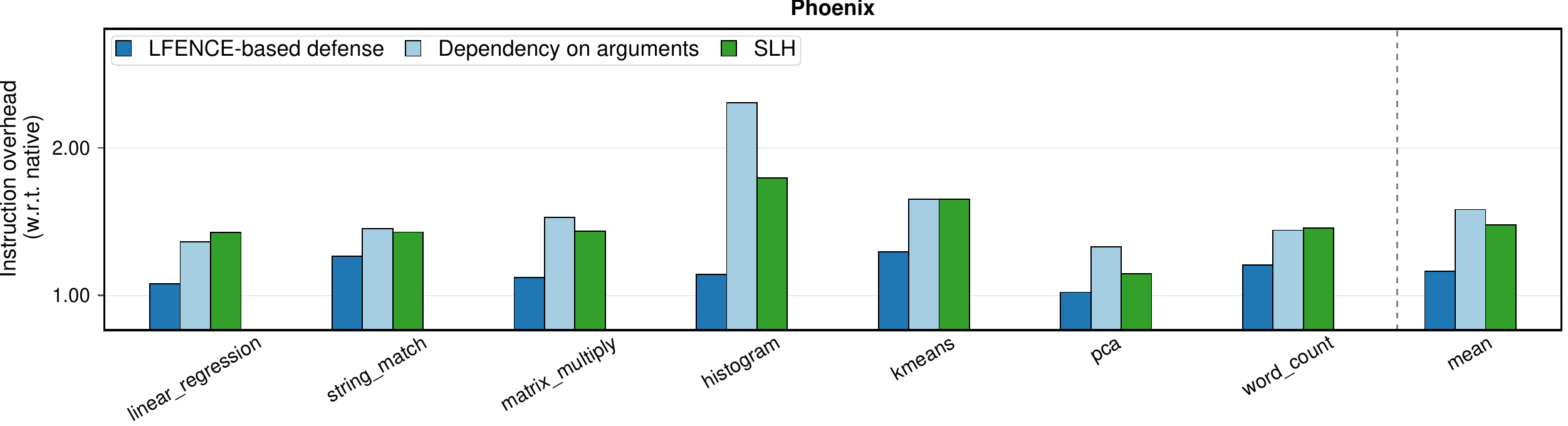}
    \caption{Increase in number of instructions with respect to native version. (Lower is better.)}
    \label{fig:instr}
\end{figure*}

\section{Preventing speculation}
\label{sec:preventing}

The most straightforward way of defending against \bcb{} is to prevent the speculation itself.
Consider the example in \figref{prevent}a:
illegal memory access will never happen if the read (line 9) runs strictly after the comparison (line 4).

There are two alternative approaches to enforcing this property: completely preventing speculation via serialization instructions, and delaying loads by adding artificial data dependencies.

\subsection{Serialization}

Intel documentation~\cite{Intel18a} proposes to patch vulnerable regions of code by explicitly serializing them with an \lfence{}, an instruction that ensures that all prior instructions execute before it, and all later instructions---after\footnotemark[1].
We can force the load to wait for the comparison by adding an \lfence{} in-between (see \figref{prevent}b).
To protect the application entirely, we would have to add an \lfence{} after every comparison\footnotemark[2] or, more precisely, after every conditional branch.

\footnotetext[1]{Before the publication of Spectre, the documentation described \lfence{} as an instruction that only prevent reordering of loads and not other instructions.
Afterward, Intel uncovered that \lfence{} is, in fact, a full serialization instruction~\cite{Intel18a,Intel18b}.}

\footnotetext[2]{Mind that we use comparison only as an example.
In practice, any operation that modifies \code{EFLAGS} could be used as a branch condition.}

The approach is, however, excessive because it delays \emph{all} the instructions after the comparison, not only the vulnerable load.
In \figref{prevent}b, lines 8 and 12 do not access memory and can safely run in parallel with the comparison.
And yet, the \lfence{} delays them too.

\subsection{Artificial data dependencies}

A more efficient approach is to allow most of the instructions to benefit from speculation and delay only those that read from memory (we assume that any data in memory could be security sensitive).

Modern CPUs already have a mechanism for ensuring that one instruction runs strictly after another without delaying the rest of the instructions: data dependency.
The approach is to reuse this mechanism by adding an artificial data dependency between conditional jumps and later loads.
In the previous example (\figref{prevent}a), if we add a data dependency between the comparison (line 4) and the loads (lines 9 and 13), the load will be delayed while the benign operations on lines 8 and 12 can still benefit from parallelism.

\section{Ways of introducing a data dependency}
\label{sec:dependency}

The idea behind the dependency-based approaches is to delay all instructions using the secret until the comparison is resolved by masking the secret with a value data-dependent on \code{EFLAGS}.
There are two ways to get such value: either by reading \code{EFLAGS} directly (\code{LAHF} instruction) or by using a conditional move.
Alternatively, the secret could be masked with comparison arguments, although it provides much weaker ordering guarantees.

\subsection{Dependency via LAHF}

The simplest way to introduce the dependency is to use \code{LAHF}, an instruction that stores the value of \code{EFLAGS} into \code{RAX} (\figref{prevent}c, line 5).
We could reserve a register (e.g., \code{R15}) and modify it using the stored flags (e.g., via \code{XOR}) to create a data dependency (line 6).
Later, we twice \code{XOR} the secret with \code{R15} (lines 10--11) thus making all further instructions using the secret dependent on the comparison, but without actually changing the secret's value.

The main issue of this approach is that we have to temporary store (line 3) and restore (line 7) the value of \code{RAX} every time we invoke \code{LAHF}.
We cannot reserve \code{RAX} as we did with \code{R15} because many instructions rely on this register.
Correspondingly, it increases the runtime cost of the protection.

\subsection{Dependency via conditional move}

To avoid the cost of keeping the \code{RAX} state, we could use a conditional move, which is the approach used by Speculative Load Hardening (SLH)~\cite{SLH}.

SLH creates the dependency via \code{CMOV}, an instruction that performs a move based on the value of one of the status flags in the \code{EFLAGS} register.
In \figref{prevent}d, the secret is masked (line 10) with a value that may be set to zero (line 5) if the comparison and conditional move mismatch (i.e., if we have a misprediction).
It has a double effect.
First, similarly to \code{LAHF}-based defence, SLH makes the loads data dependent on the comparison, which prevents the speculation.
Second, SLH zeroes out the loaded value (lines 5 and 10) in case of misspeculation.
Although it is redundant on current hardware, future generations of Intel CPUs may introduce a value prediction feature that can speculate even in the presence of data dependency.

Since the mask could have only one of the two values---either all ones or zero---there is no need to make a double \code{XOR} and a single \code{AND} is sufficient (line 10).

\subsection{Dependency on arguments}

The approach used by SLH could be simplified even further.
Instead of creating a dependency on \code{EFLAGS}, we could add a dependency on the comparison arguments (see \figref{prevent}e, line 3).
Hence, the comparison can run in parallel with the loads, while the dependency ensures that the leaky load will start only when the arguments are either in registers or in L1 cache.
In this case, the speculation window will likely last only 1--2 cycles.

Although this approach may prevent the leak in many cases, it does not provide any strict guarantees of ordering.
If the CPU reorders the instructions such that the comparison begins \emph{after} the loads (e.g., because of an internal hardware hazard), the attack can still succeed.
If the attacker comes up with a way to delay the comparison reliably, it will render this strategy ineffective.


\section{Evaluation}
\label{sec:evaluation}

In this section, we evaluate the performance impact of the approaches descussed in \secref{preventing} and \secref{dependency}.
We used the author's implementation of SLH and we implemented the other two approaches on our own.

\myparagraph{Experimental Setup}
The experiments were carried out on a machine with a 4-core Intel processor operating at 3.3~GHz (Haswell microarchitecture) with 32GB of RAM, a 256GB SATA-based SDD, and running Linux kernel 4.15.
Each core has private 32KB L1 and 256KB L2 caches, and all cores share an 8MB L3 cache.
We used the largest available datasets provided by the Phoenix benchmark suite.
As of compilers, we used LLVM 7.0 for SLH and LLVM 5.0 for the other two approaches.

The numbers are normalized against the native LLVM of the corresponding version.
For all measurements, we report the average over ten runs and geometric mean for the ``gmean'' across benchmarks.

\myparagraph{Performance}
\figref{perf} shows performance overheads of the \lfence{}-based defense, SLH, and dependency on arguments, measured across the Phoenix benchmarks~\cite{Phoenix}.

As we see, adding \lfence{}s after every conditional branch is extremely expensive and causes 440\% slowdown on average.
Such a high overhead appears because \lfence{} virtually disables speculative execution.
As \figref{ipc} shows, the application cannot use the available instruction-level parallelism to its full extent:
With the \lfence{}s, the average number of instructions per cycle (IPC) drops from \textasciitilde2.3 to \textasciitilde0.5.
As of the dependency on arguments and SLH, they delay only memory accesses and therefore, IPC drops only to \textasciitilde2.

IPC is not the only influencial factor, though.
\emph{histogram} has lower overhead with SLH, yet the IPC is lower than with the dependency on arguments.
Here, the overhead is mainly caused by additional instructions (see \figref{instr}).
SLH uses only a single \code{AND} for masking, whereas the other approaches need two \code{XOR}s.
\emph{Histogram}, having more loads and fewer loops than other benchmarks, has this effect more pronounced.

As of the extreme cases, \emph{pca} contains large loops with many arithmetic operations (mainly vectorized) on its hot path, hence the speculation has less influence on it.
On the other side of the spectrum are \emph{kmeans} and \emph{string\_match}.
Here, the high overheads are caused by tight loops on the hot path.
Both SLH and the dependency on arguments force the loops to run sequentially thus reducing the level of parallelism (see \figref{ipc}).
SLH versions are slower because SLH uses a more expensive instruction to instrument conditional branches (\code{CMOV} instead of \code{XOR}).

%

\section{Conclusion}
\label{sec:conclusion}

We presented an overview of possible approaches to preventing Bounds Check Bypass by creating artificial data dependencies between conditional jumps and subsequent memory loads.
Because of allowing benign instructions to run in parallel with the jumps, these approaches achieve much better utilization of the available CPU resources in comparison to serialization with \lfence{}s.
In our experiments, they introduce 60\% overhead, while \lfence{}-based defense causes 440\% slowdown.

\balance

{\footnotesize \bibliographystyle{acm}
\bibliography{refs.bib}


\end{document}